\documentclass[aps,prb,amsfonts,amssymb,twocolumn,amsmath,preprintnumbers,nofootinbib,floatfix,showpacs]{revtex4}
\usepackage{amsmath,bm,amsfonts,amssymb}
\usepackage[dvips]{graphics}
\usepackage{graphicx}

\begin{document}

\title{Influence of the atomic scale inhomogeneity of the pair interaction on the local pair formation and density of states
in high-$T_c$ superconductors.}
\author{A. M. Bobkov}
\affiliation{Institute of Solid State Physics, Chernogolovka,
Moscow reg., 142432 Russia}
\author{I. V. Bobkova}
\email[Electronic address: ]{bobkova@issp.ac.ru}
\affiliation{Institute of Solid State Physics, Chernogolovka,
Moscow reg., 142432 Russia}

\date{\today}

\begin{abstract}
The influence of the atomic-scale inhomogeneities of the pairing
interaction on the superconducting order parameter distribution
and the LDOS is studied in the framework of mean-field BCS theory
for two-dimensional lattice model. It is found that the ratio of
the local low-temperature gap in differential conductance to the
local temperature of vanishing the gap $2\Delta_g/T_p$ can take
large enough values compared to the homogeneous case. This ratio
practically does not depend on the location in the sample and is
independent on the concentration of local pair interaction
perturbations in wide range of concentrations. The obtained
results could bear a relation to the recent measurements by Gomes
{\it et. al}\cite{gomes07}.
\end{abstract}

\maketitle

Nanoscale inhomogeneities have been widely observed in the
high-temperature superconductor $\rm Bi_2Sr_2CaCu_2O_{8+x}$
(BSCCO) and have generated intense interest
\cite{cren00,pan01,howard01,lang02,mcelroy05,fang06}. In
particular, the spectral gap in the local density of states (LDOS)
has been investigated by scanning tunneling microscopy (STM). It
was found that the gap varies by a factor of 2 over distances of
$20-30 $ $\overset{\circ}{\rm A}$. These observations have been
made primarily in the superconducting state. Several scenarios
have been proposed to understanding this electronic inhomogeneity.
First of all, it was speculated that poorly screened electrostatic
potentials of the dopant atoms vary a doping concentration
locally, giving rise to the gap modulations
\cite{martin01,zwang02,qwang02}. Alternatively, these
inhomogeneities are associated with a competing order parameter
\cite{kivelson03,atkinson05,alvarez05,ghosal04,podolsky03,chen04}.
Further, the positive correlations between the inhomogeneities and
positions of the dopant atoms have been observed by STM on the
optimally doped BSCCO \cite{mcelroy05}. After that it was proposed
by Nunner {\it et al.} \cite{nunner05} that the dopant atoms
modulate the pairing interaction locally on the atomic scale. The
LDOS calculated in the framework of this model is in good
agreement with the key characteristics of the experimental data. 
In addition, finite-temperature order parameter evolution 
was investigated and the specific heat was calculated within the same
model\cite{andersen06}. 
 
On the other hand, it is well known that in the high-$T_c$
superconductors a partial gap in the LDOS exists for a range of
temperatures above $T_c$ \cite{timusk99}. There is no consensus up
to now if this gap is due to pairing without phase coherence, a
competing order or proximity to the Mott state
\cite{norman05,lee06,millis06,cho06}. The inhomogeneities
described above complicate the situation. Only very recently the
spatially resolved STM measurements of gap formation in BSCCO
samples were performed \cite{gomes07}. For a range of doping from
$0.16$ to $0.22$ they have found that gaps nucleate in nanoscale
regions above $T_c$ and proliferate as the temperature is lowered,
evolving to the spatial distribution of gap values in the
superconducting state. It was observed experimentally that
overdoped and optimally doped samples have identical
gap-temperature scaling ratios, which together with the fact that
in the overdoped samples pseudogap effects are believed to be weak
or absent, allowed Gomes $\it et. al$ to interpret the gaps above
$T_c$ as those associated with pairing. The most striking
experimental observation is that, despite the inhomogeneity, every
pairing gap develops locally at the temperature $T_p$ following
the relation $2\Delta/T_p=7.9 \pm 0.5$ in wide range of doping
from overdoped to optimally doped samples. So large and
independent on the size of local gap values ratio $2\Delta/T_p$
seems to be different from the expectations based on the BCS
theory and its strong-coupling extension, where it is in the range
$3.5-5$ and becomes dependent on $\Delta$.

In the present paper we show that if the pairing interaction is
modulated locally on the atomic scale, as it was proposed by
Nunner {\it et. al} \cite{nunner05}, the ratio $2\Delta/T_p$
strongly increases in the framework of conventional BCS theory.
This is in sharp contrast to the influence of the potential
disorder, which, as was demonstrated \cite{franz97}, can only
diminish this quantity. It is found that the ratio $2\Delta/T_p$
is practically independent on the number of off-diagonal (pairing
interaction) scatterers, which is proportional to the doping in
the model by Nunner {\it et. al} \cite{nunner05}, in a wide range
of concentrations. We obtained the ratio $2\Delta/T_p$ to be
position-independent in case if the off-diagonal scatterers are
distributed rather uniformly and equal to $5.9$. For comparison,
the ratio $2\Delta/T_p=4.7$ for the homogeneous sample
corresponding to the lattice parameters we use (it is slightly
higher than the ratio $2\Delta/T_c=4.3$, because $T_p$ is smaller
than the mean-field critical temperature $T_c$ as it is discussed
below). These results are reminiscent of the measurements by Gomes
{\it et. al}\cite{gomes07} except for the fact that the ratio
$2\Delta/T_p$ we calculated is about thirty percent smaller than
experimentally measured one. On the other hand, the ratio
$2\Delta_g/T_p$ is found to be very sensitive to characteristic
size and height of the individual pairing interaction scatterer
and especially to the choice of the lattice parameters and can be
further increased manipulating by these quantities. It is worth
noting that our analysis is phenomenological and the conclusions
are independent on the underlying pairing mechanism and the
particular course of the local pair interaction modulations.

{\it Model and method.$-$}We consider the following mean-field
Hamiltonian on a square lattice
\begin{equation}
\hat H \!=\! -\!\!\sum \limits_{\langle ij \rangle, \sigma}\!
t_{ij} c_{i\sigma}^\dagger c_{j \sigma}-\!\sum
\limits_{i,\sigma}\mu c_{i\sigma}^\dagger c_{j \sigma}+\!\sum
\limits_{\langle ij \rangle}\!\left(\!
\Delta_{ij}c_{i\uparrow}^\dagger c_{j\downarrow}^\dagger \!+ \!
h.c.\! \right), \label{hamiltonian}
\end{equation}
where $c_{i\sigma}(c_{i\sigma}^\dagger)$ stands for an electron
annihilation (creation) operator at site $i$ with spin $\sigma$.
$\sum_{\langle ij \rangle}$ indicates summation over neighboring
sites, $t_{ij}$ is the hopping integral between sites $i$ and $j$.
We set $t_{ij}$ to be $t=1$ for the nearest-neighbor hopping,
$t^\prime=-0.3$ for the next nearest-neighbor hopping and
$t^{\prime \prime}=0.1$ for the next-next one. $\mu$ is adjusted
to be $-1$ to model the Fermi surface of BSCCO near optimal
doping. Below all the energies are measured in units of $t$. The
nearest-neighbor $d$-wave order parameter (OP) should be
determined self-consistently: $\Delta_{ij}=-g_{ij}\langle
c_{i\downarrow}c_{j\uparrow}-c_{j\downarrow}c_{i\uparrow}
\rangle$. Following Ref.~\onlinecite{nunner05} we model the
inhomogeneous pairing interaction by $g_{ij}=g_b+\delta
g(f_i+f_j)/2$, where $g_b$ corresponds to an average background
interaction and $f_i=\sum_s \exp(-r_{is}/\lambda)/r_{is}$, where
$r_{is}$ is the distance between the site $i$ and the source of
the pairing interaction perturbation. The mean-field BCS treatment
is generally believed to be appropriate for overdoped (and, to
some extent, optimal doped) samples. So, we do not consider
underdoped regime, where the proximity to the Mott state should be
taken into account.

In order to analyze the inhomogeneous pairing correlations in the
framework of the mean-field hamiltonian (\ref{hamiltonian}) we
exploit the fully self-consistent T-matrix technique for Gor'kov
Green's functions. The full Green's function takes the form
\begin{equation}
\check G_{ij}=\check G_{ij}^0 + \sum \limits_{k,m} \check G_{ik}^0
\check T_{km} \check G_{mj}^0 \label{Green_function} \enspace .
\end{equation}
Here $\check T_{km}=-\sum \limits_n (\check M^{-1})_{kn}\check
V_{nm}$, $\check M_{km}=\delta_{km} + \sum \limits_n \check
G_{kn}^0 \check V_{nm}$. In this paper we only focus on the
off-diagonal self-energy inhomogeneity and therefore $\check
V_{km}=\delta \Delta_{km} i \hat \sigma_2 i \hat \tau_2$. All
Green's functions and T-matrices are $4 \times 4$ matrices in the
direct product of spin and particle-hole spaces, what indicated by
the symbol $\check ~$. $\hat \tau_i$ and $\hat \sigma_i$ are Pauli
matrices in particle-hole and spin spaces respectively. The
summation is taken over all the sites, where the OP
$\Delta_{km}=\Delta_{km}^0+\delta \Delta_{km}$ differs from the
background value $\Delta_{km}^0$. $\Delta_{km}^0$ is assumed to be
of $d$-wave type, that is $\Delta_{ii \pm \hat a}^0=-\Delta_{ii
\pm \hat b}^0=\Delta^0$. $\hat a$ and $\hat b $ are basis vectors
of the square lattice. We set the lattice constant $a$ to be equal
to unity.

{\it Local order parameter.$-$}OP is to be calculated from the
self-consistency equation
\begin{equation}
\Delta_{ij} = g_{ij}T\sum \limits_{\varepsilon_n}{\rm Tr}_4 \left[
\hat \tau_- i \hat \sigma_y \check G_{ij}(\varepsilon_n) \right]
\label{self_consistency} \enspace ,
\end{equation}
where $\hat \tau_-=(\hat \tau_x - i \hat \tau_y)/2$. Eqs.
(\ref{self_consistency}) and (\ref{Green_function}) allow us to
find the OP $\Delta_{ij}$ numerically.

We begin by considering a single perturbation of the pairing
interaction of the form $f_i=\exp (-r/\lambda)/r$, where
$r=\sqrt{i_x^2+i_y^2+z^2}$. However, the results do not depend
qualitatively on the particular form of the perturbation and are
only controlled by its effective width and height.

\begin{figure}[!tbh]
\begin{minipage}[b]{.5\linewidth}
     \centerline{\includegraphics[clip=true,width=1.7in]{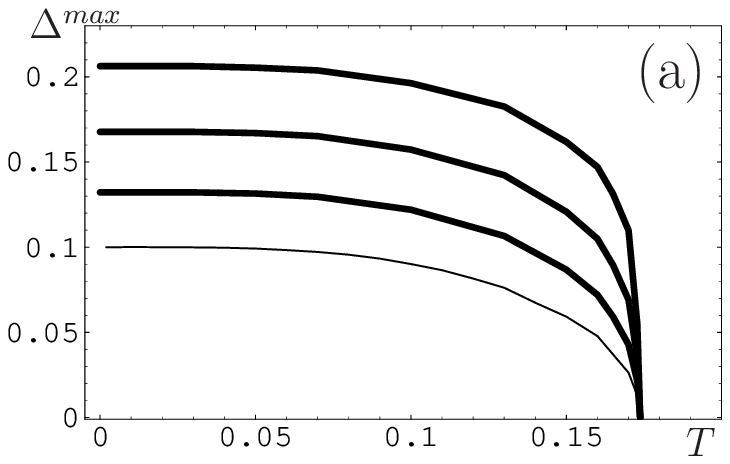}}
     \end{minipage}\hfill
     \begin{minipage}[b]{.5\linewidth}
   \centerline{\includegraphics[clip=true,width=1.7in]{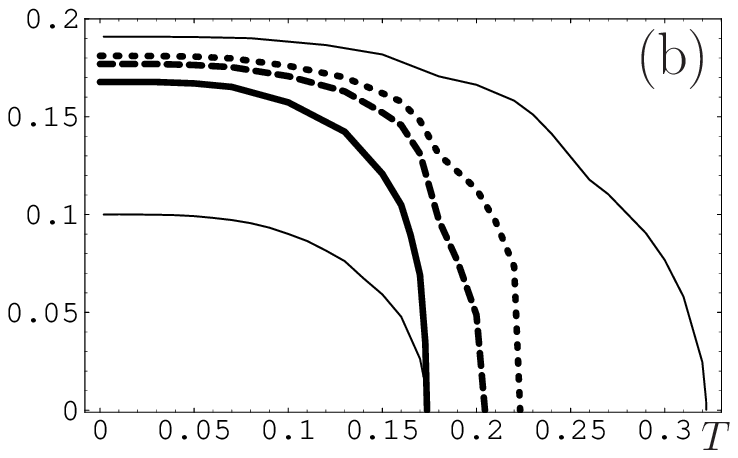}}
  \end{minipage}
        \caption{(a) Dependence of the site-averaged OP maximal
value $\Delta^{max}$ on temperature for $z=0.5$ and $\lambda=0.5$.
$g^{max}=g_0+\delta g^{max}=0.857$, $1.022$ and $1.187$ from the
bottom curve to the top one.
       (b) Evolution of $\Delta^{max}(T)$ dependence with increasing of the
       perturbation
size for constant $g^{max}=1.022$. $z=0.5$, $\lambda=0.5$ (solid
line); $z=2$, $\lambda=2$ (dashed line) and $z=3$, $\lambda=3$
(dotted line). Bottom and top thin solid lines represent
homogeneous bulk dependence $\Delta_0(T)$ for $g_b=0.692$ and
$g_b=1.022$, respectively.} \label{Delta_T_single}
\end{figure}

The dependence of the site-averaged OP maximal value
$\Delta^{max}=(\Delta_{0,\hat a}+\Delta_{0,-\hat a}+|\Delta_{0,
\hat b}|+|\Delta_{0,-\hat b}|)/4$ on temperature is presented in
Fig.\ref{Delta_T_single}. The left panel demonstrates the most
spikier perturbation we consider: $z=0.5$ and $\lambda=0.5$. This
perturbation has considerably non-zero value only at four bonds
emanating from the central site. Here and below all the distances
are measured in units of lattice constant. The different curves
correspond to different heights of the perturbation:
$g^{max}=g_b+\delta g^{max}=0.857$, $1.022$ and $1.187$ from the
bottom curve to the top one. The background pairing interaction is
taken to be $g_b=0.692$. The thin solid line represents the
temperature behavior of the homogeneous order parameter without a
perturbation added. We denote the critical temperature of the
inhomogeneous sample by $T_x$ in order to distinguish it from the
background critical temperature $T_b$. If the perturbation size is
very small (a few bonds), as in Fig.~\ref{Delta_T_single}(a), the
corresponding $T_x$ practically does not differ from $T_b$. The
physical reason for it is that small as compared to
superconducting coherence length perturbations at the mean-field
level cannot maintain superconductivity by themselves and only do
this due to the superconductivity in the bulk. Then the bulk OP
vanishes, the pairing correlations in the small area go to zero
abruptly. On the other hand, the value of the zero-temperature OP
$\Delta^{max}$ strongly enhances when the height of the
perturbation grows. Therefore, local ratios $\Delta_{ij}/T_x$ can
considerably exceed the homogeneous bulk value. For the parameters
we consider in Fig.~\ref{Delta_T_single}(a) $\Delta_0/T_b=0.58$
and $\Delta^{max}/T_x=0.76$, $0.97$ and $1.19$ from bottom to top.

When the size of the pairing interaction perturbation increases,
$T_x$ grows and, correspondingly, the ratio $\Delta_{ij}/T_x$
starts to decline. Naturally, when the size of the enhanced
pairing area becomes of the order of a few superconducting
coherence lengths $\xi_s$, the bulk value of $\Delta_{ij}/T_x$
corresponding to $g=g_b+\delta g$ should be restored at the center
of the cluster. The evolution of $\Delta^{max}(T)$ dependence with
increasing of the perturbation size is illustrated in
Fig.~\ref{Delta_T_single}(b) for constant $g^{max}=1.022$. For
comparison the curve $\Delta_0(T)$ corresponding to this value of
the pairing interaction is depicted in the same panel by the thin
line. Therefore, large enough ratios $\Delta^{max}/T_x$ could be
only obtained when the characteristic size of the OP enhanced area
is considerably less than superconducting coherence length.
However, the anomalous Green's function, which enters the
self-consistensy equation (\ref{self_consistency}) always has the
characteristic length of the order of $\xi_s \sim  t/\Delta_0$
regardless of the particular parameter of the hamiltonian coursing
the inhomogeneity. Consequently, the OP spacial profile follows
the spacial profile of the coupling constant according to
Eq.~(\ref{self_consistency}) except for very low tails extended to
the distance of the order $\xi_s$ due to the inhomogeneities of
the anomalous Green's function. Therefore, we believe that the
large enough ratio $\Delta_{ij}/T_x$ can only be obtained (at least in the
framework of mean-field BCS theory with temperature-independent
coupling constant) by assuming the atomic-scale inhomogeneity of
the pairing interaction strength.

Now we turn to discussion of more realistic situation, when many
pair interaction scatterers are present in the sample. It is worth
noting that in our mean-field treatment we neglect OP phase, which
is a strongly fluctuating quantity in short coherence length
high-$T_c$ superconductors and, especially, in the inhomogeneous
situation. In the present paper we only study local pairing and do
not concern the global transition temperature $T_c$, which is
controlled by the phase fluctuations.

\begin{figure}[!tbh]
\begin{minipage}[b]{.5\linewidth}
   \centerline{\includegraphics[clip=true,width=1.7in]{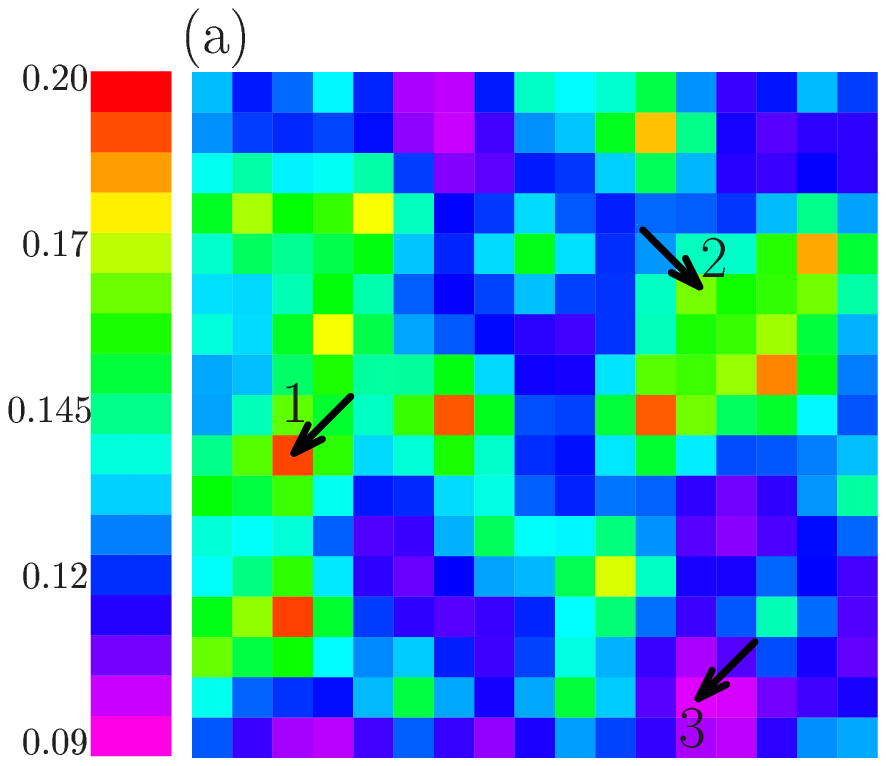}}
  \end{minipage}\hfill
 \begin{minipage}[b]{.5\linewidth}
   \centerline{\includegraphics[clip=true,width=1.7in]{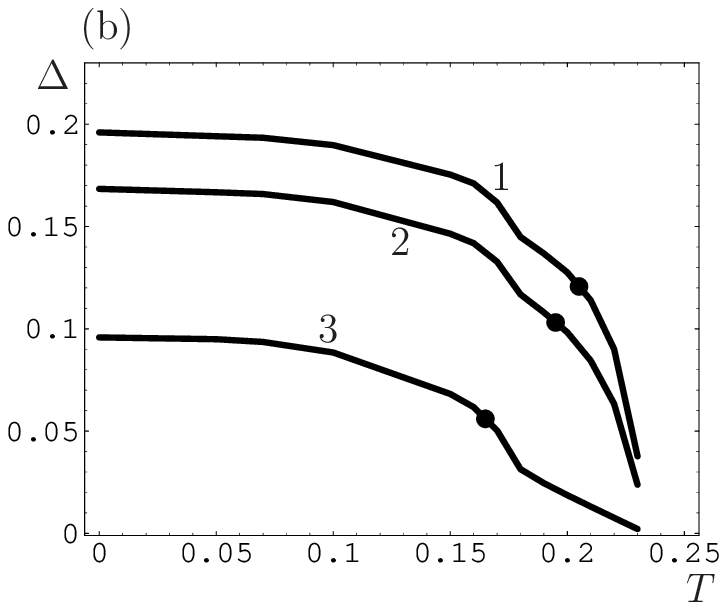}}
  \end{minipage}
  \begin{minipage}[b]{.5\linewidth}
   \centerline{\includegraphics[clip=true,width=1.7in]{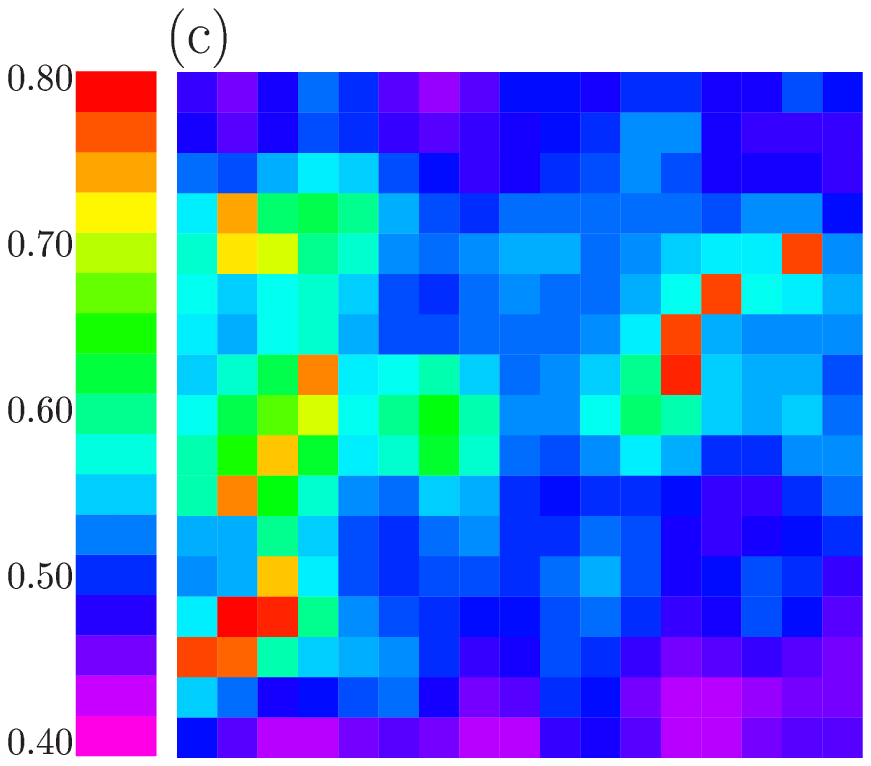}}
  \end{minipage}\hfill
 \begin{minipage}[b]{.5\linewidth}
   \centerline{\includegraphics[clip=true,width=1.7in]{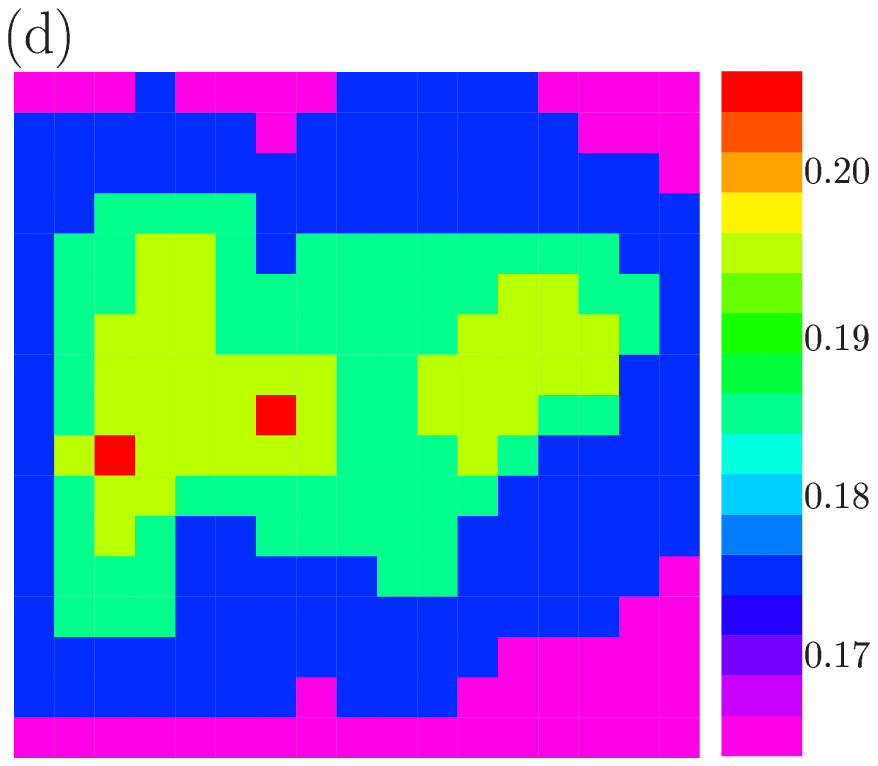}}
  \end{minipage}
  \caption{(Color online) (a) Zero-temperature distribution of the site-averaged
  OP.
  (b) The temperature dependence of the site-averaged OP for the
  sites marked by the arrows in (a). The bold circles show $T_p$ correspondinding to
  these locations.
  (c) Low-temperature gap map. (d) $T_p-$map.} \label{Delta_T_many}
\end{figure}

We have considered a $21 \times 21$ sites square as a perturbation
described by the T-matrix. The coupling constant outside this
square is set to be $g_b=0.692$. Fig.~\ref{Delta_T_many} shows the
particular example. The pairing interaction scatterers
corresponding to $\lambda=1.5$ and $z=1$ are distributed in the
square with the concentration $n=0.078$. The background coupling
constant in the square and the height of the individual
perturbation are chosen to give the same average as outside the
square: $g_0=0.45$, $\delta g$ is randomly distributed in the
range $0.70-1.35$.

The resulting distribution of zero-temperature OP is presented in
Fig.~\ref{Delta_T_many}(a). As it was described earlier for the
single perturbation, the spacial profile of the OP inhomogeneity
mainly follows that one of the coupling constant. The dependence
of the site-averaged $\Delta_{i}(T)$ for the marked sites is
plotted in Fig.~\ref{Delta_T_many}(b).

{\it Local density of states.$-$}STM technique measures the local
differential conductance, and the thermally smeared LDOS,
described by the expression
\begin{equation}
dI/dV=\int \limits_{-\infty}^{\infty} d\varepsilon
(df(\varepsilon+V)/dV)\rho_i(\varepsilon,T) \label{smeared_DOS}
\enspace ,
\end{equation}
can be extracted from these measurements. Here
$\rho_i(\varepsilon,T)=-(1/\pi){\rm Im}G_{ii}^R(\varepsilon,T)$ is
the local density of states, $f(\varepsilon)$ is Fermi
distribution function and $V$ is voltage applied between the STM
tip and the sample. For the homogeneous situation in the framework
of mean-field weak-coupling BCS theory the distance between the
coherence peaks $2\Delta_g$ in the conductance spectra directly
connected to the OP by the simple relation $\Delta_g \approx 3.7
\Delta(T=0)$ for the particular
lattice parameters we have chosen.

For an inhomogeneous system, where the characteristic size of the
patch $\lesssim \xi$, there is no any direct simple relationship
between the local OP and the local gap $\Delta_g$. One can only
conclude from the numerical calculations \cite{nunner05,fang06}
that for regions, where the OP is enhanced from the background,
the gap gets wider and the peak height is suppressed compared to
the average value. It is worth noting that, unlike the homogeneous
situation, this peak does not represent the maximal
superconducting gap on the Fermi surface, but rather originates
from the spectral weight transfer from the nearby van Hove
singularity due to the Andreev scattering processes. Otherwise, if
the OP in a cluster is less than that one in the background, the
narrow and high Andreev resonant peaks develop in the cluster
region resulting in diminishing of the gap region. For this reason
we investigate not only the OP distribution, but also the
experimentally measurable thermally smeared LDOS, which has a
maximum at $V=\Delta_g$. The corresponding low-temperature gap map
is represented in Fig.~\ref{Delta_T_many}(c) for the model sample
considered above.

\begin{figure}[!tbh]
\begin{minipage}[b]{.5\linewidth}
   \centerline{\includegraphics[clip=true,width=1.7in]{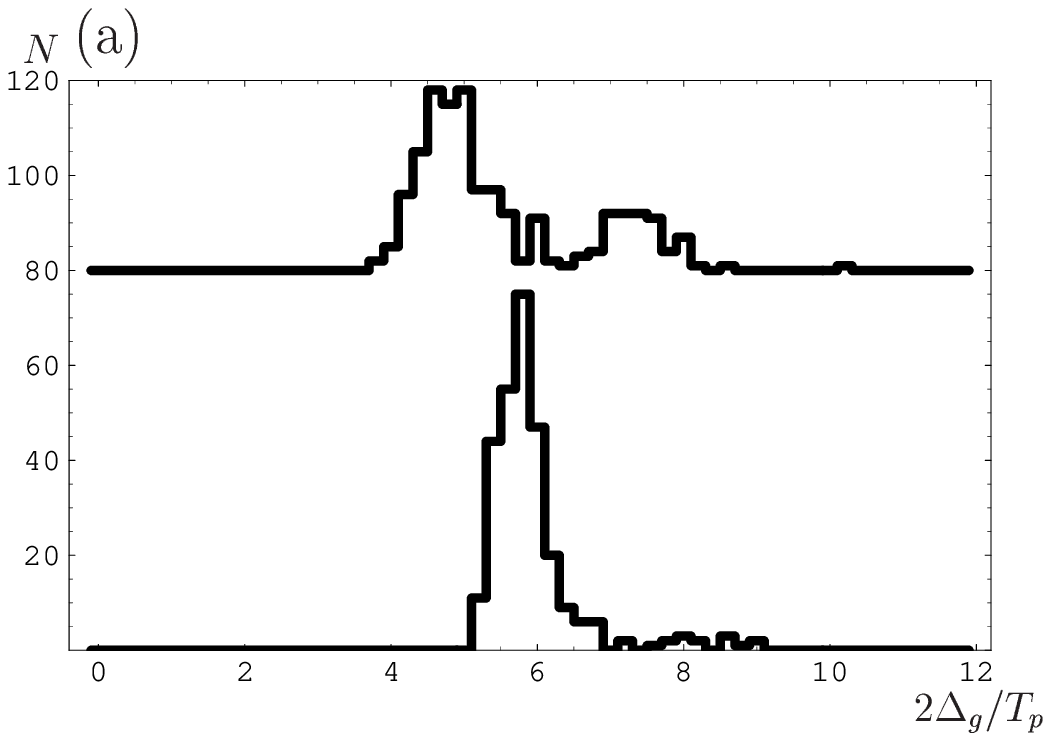}}
  \end{minipage}\hfill
 \begin{minipage}[b]{.5\linewidth}
   \centerline{\includegraphics[clip=true,width=1.7in]{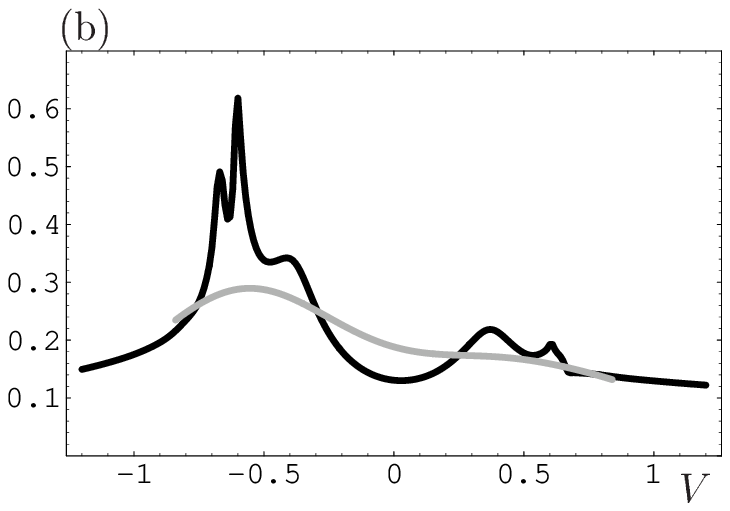}}
  \end{minipage}
    \caption{(a) The probability distribution to find the particular value for
$2\Delta_g(i)/T_p(i)$. The bottom curve represents the same model
sample as in Fig. \ref{Delta_T_many}. The top one corresponds to
another example with $n=0.052$, $\lambda=2$, $z=0.5$, $g_0=0.35$
and $\delta g=0.8$. The offset is for clarity. (b) LDOS (black
curve) in comparison with the thermally smeared LDOS (gray curve)
for a particular location and $T=0.18$.} \label{ratio}
\end{figure}

Experimentally \cite{gomes07} the temperature $T_p(i)$ of the gap
disappearing for the particular location in the sample has been
determined using the criterion $dI/dV(V=0) \ge dI/dV$ (for all $V
\ge 0$). Using the above criterion we calculated the distribution
$T_p(i)$ from the thermally smeared LDOS curves. The corresponding
$T_p$-map is represented in Fig.~\ref{Delta_T_many}(d). In
addition, for the locations, marked in Fig.~\ref{Delta_T_many}(a),
appropriate $T_p$ is shown by bold circles in
Fig.~\ref{Delta_T_many}(b). Although in our model the OP vanishes
at the temperature $T_x$, which is the same for the entire sample,
it is seen that the temperature $T_p$ is strongly
position-dependent and considerably lower than $T_x$ for the most
part of the sample. There are two main physical reasons for this
effect: (i) LDOS is essentially nonlocal (on the atomic scale)
quantity with the characteristic size of order of $\xi_s$. This
fact leads to the partial averaging of essentially different gaps
over the region $\sim \xi_s^2$ and spacial redistribution of the
spectral weight due to Andreev scattering processes. (ii) thermal
broadening of the LDOS further washes out the conductance peaks.
It is worth noting that the LDOS $\rho(\varepsilon, T)$ taking
without thermal smearing results in higher values of the local gap
vanishing temperature. This is demonstrated in the right panel of
Fig.~\ref{ratio}.

The probability distribution to find the particular value for
$2\Delta_g(i)/T_p(i)$ in the model sample considered above is
plotted in the left panel of Fig.~\ref{ratio} (bottom curve). It
is seen that the distribution is quite narrow, that is this ratio
is practically independent on the particular location in the
sample. In the framework of our model this result can be only
obtained  if the scatterers are distributed in the sample
randomly, but the additional restriction is imposed: they
influence each other in the sense that the distance between them
cannot be too small. If this restriction is not imposed and the scatterers
are distributed quite randomly, the ratio $2\Delta_g(i)/T_p(i)$ is
obtained to be more position-dependent. This is demonstrated for
another model sample in the left panel of Fig.~\ref{ratio} (top
curve). This sample differs from the considered above by the shape
of an individual scatterer and by their concentration. However, we
have investigated a number of model samples and checked that the
$2\Delta_g(i)/T_p(i)$ distribution shape is only determined by the
extent of uniformity of the scatterers.

The averaged ratio $2\Delta_g/T_p$ we calculated is equal to
$5.9$. It is larger than that one for a homogeneous situation
corresponding to the same normal state hamiltonian
$2\Delta_g/T_p=4.7$ but is smaller than
experimentally measured one\cite{gomes07}. However, it is obvious
from the above analysis that the underlying ratio of the OP to the
critical temperature $\Delta_{ij}/T_x$ gets larger as the height
of the perturbation increases and its width diminishes. The
manifestation of the effect in LDOS, and, consequently, the
behavior of the experimentally measured ratio $2\Delta_g/T_p$ is
not so straightforward\cite{bobkovy} and strongly depends on the
particular choice of the lattice parameters. The point is that for
the inhomogeneous situation the peak in the LDOS results from the
Andreev scattering processes between the background
superconducting coherence peak and nearby van Hove singularity, as
it was already mentioned above. So, the position $\Delta_g$ of the
LDOS peak strongly depends on the energy distance between them in
comparison to the strength of the individual scatterer.

In conclusion, we have studied the influence of the atomic-scale
inhomogeneities of the pairing interaction strength on the OP
distribution and the LDOS in the framework of mean-field BCS
theory. It is found that the ratio of the local low-temperature
gap in differential conductance spectra to the local temperature
of vanishing the gap $2\Delta_g/T_p$ can take large enough values
compared to the homogeneous one. This ratio is
position-independent in case if the off-diagonal scatterers are
distributed rather uniformly and practically independent on their
concentration in wide range of the concentrations. On the other
hand, it is quite sensitive to characteristic size and height of
the individual pairing interaction scatterer and to the choice of
the lattice parameters.

%\begin{figure*}[!tbh]
%\begin{minipage}[b]{.5\linewidth}
 %  \centerline{\includegraphics[clip=true,width=2in]{fig2a_left.eps}}
  %\end{minipage}\hfill
 %\begin{minipage}[b]{.5\linewidth}
  % \centerline{\includegraphics[clip=true,width=2in]{fig2a_right.eps}}
  %\end{minipage}
  %\begin{minipage}[b]{.5\linewidth}
   %\centerline{\includegraphics[clip=true,width=2in]{fig2b_left.eps}}
 % \end{minipage}\hfill
 %\begin{minipage}[b]{.5\linewidth}
  % \centerline{\includegraphics[clip=true,width=2in]{fig2b_right.eps}}
  %\end{minipage}
  %\begin{minipage}[b]{.5\linewidth}
   %\centerline{\includegraphics[clip=true,width=2in]{fig2c_left.eps}}
  %\end{minipage}\hfill
 %\begin{minipage}[b]{.5\linewidth}
  % \centerline{\includegraphics[clip=true,width=2in]{fig2c_right.eps}}
  %\end{minipage}
%\caption{ } \label{Keldysh}
%\end{figure*}

{\it Acknowledgments.} The support by RFBR Grant 05-02-17731
(A.M.B.) and the programs of Physical Science Division of RAS is
acknowledged. I.V.B. was also supported by the Russian Science
Support Foundation and RF Presidential Grant No.MK-4605.2007.2
%\bibliography{/users/tkm/howell/latexx/bibtexx/refs}

\end{document}